\documentclass[12pt,preprint]{aastex}

\shorttitle{Detection of HC$_4$N in IRC+10216}
\shortauthors{Cernicharo et al.}

\begin{document}

\title{Detection of the linear radical HC$_4$N in IRC+10216}

\author{J. Cernicharo\altaffilmark{1}, M. Gu\'elin\altaffilmark{2} and
J.R. Pardo\altaffilmark{1}}
\email{cerni@damir.iem.csic.es, guelin@iram.fr, pardo@damir.iem.csic.es}

\altaffiltext{1}{Dpt. Molecular and Infrared Astrophysics (DAMIR).
IEM-CSIC. C/Serrano 121. 28006 Madrid. Spain}
\altaffiltext{2}{IRAM. 300 rue de la Piscine. F-38406 St. Martin d'H\`eres.
France}

\begin{abstract}
We report the detection of the linear radical HC$_4$N in the 
C-rich envelope of IRC+10216.
After HCCN \citep{Gue91},  HC$_4$N is the second member of the allenic 
chain family HC$_{2n}$N observed in space.
The column density of HC$_4$N is found to be 1.5 10$^{12}$ cm$^{-2}$. 
The abundance ratio HC$_2$N/HC$_4$N is
9, a factor of two larger than the decrement observed for the
cyanopolyynes (HC$_{2n+1}$N/HC$_{2n+3}$N). 
Linear HC$_4$N has a $^3\Sigma$ electronic ground state and 
is one of the 3 low-energy isomeric forms of this molecule.
We have searched for the bent and ringed HC$_4$N isomers, but could
only derive an upper limit to their column densities  
$\le$ 3\, 10$^{12}$  cm$^{-2}$ (at 3-$\sigma$).

\end{abstract}


\section{Introduction}

The detection of long linear carbon-chain molecules in interstellar and
circumstellar clouds has come as a surprise. Ab initio calculation show
that long linear species are
usually more energetic than their ring or ring-chain isomers and are 
observed to be less stable in the earthbound environment. Yet many linear
chains, such as the cyanopolyynes HC$_{2n+1}$N, are found to be
widespread and abundant in the cold circumstellar and
interstellar clouds. This abundance of long chains and the scarcity of rings 
in the cold UV-shielded clouds yield keys to the formation of large
molecules in interstellar space and, may be, can help us to understand
to the origin of the diffuse
interstellar bands \citep{Tul98,Mai04}. 

Carbon chain molecules can be classified into several families, depending on
the structure of their linear
backbone (acetylenic backbone with alternating single and triple carbon-carbon
bonds, or cumulenic backbone with double bonds), on their electronic 
ground state (open shell or closed shell) and on their end groups (H,
CN, CH$_2$ or CH$_3$). The two most widespread families are the polyynes
(HC$_{2n+1}$N and CH$_3$C$_{2n+1}$H) and the polyacetylenic radicals
(C$_{n}$H). The most abundant is presumably the family of 
polyacetylenes (HC$_{2n}$H), which, unfortunately are non-polar,
hence not detectable at radio wavelengths. Note, however, that these
species have been detected in the mid infrared with the Infrared Space
Observatory \citep{Cer01a,Cer01b}.

It has been proposed 
that the polyynes and carbon-chain radicals 
form directly in the gas phase through reactions of CCH 
with polyynes, polyacetylenes and/or polyacetylenic ions  
(e.g. HC$_{2n+1}$N + CCH  $\rightarrow$
HC$_{2n+3}$N + H) \citep{Her89,Che93,Mil00}. Such reactions tend to insert  
two triply bonded carbon atoms into the carbon backbone,
uncoupling the formation of chains with an odd number of
C-atoms from those with an even number, which 
could explain the alternance between high and low abundances
observed as the length of the C backbone increases. For example, in
the molecular shell around IRC+10216 and in the dark cloud
TMC1, the two main astronomical sources of carbon chain molecules, the 
abundance of the C$_n$H radicals with an {\it even} number of C-atoms is 
$\simeq30$ times 
larger than that of radicals of similar size with an 
{\it odd} number of carbon atoms. In contrast, the abundance decrement between 
successive species within the even (or odd) number of C-atoms families, 
C$_n$H/C$_{n+2}$H, is found to be only 4-6 \citep{Cer87, Gue91}. 

The same seems to apply to the chains terminated by CN: the
abundance decrement between cyanopolyynes, HC$_{2n+1}$N/HC$_{2n+3}$N, 
is between 3 and 5 (for n=1-4), whereas HCCN, the first member of the `even' 
chain family, is found to be $\simeq 200$ times less abundant than HC$_3$N 
\citep{Gue91}. So far, this similarity between polyynes and carbon chains, 
could not be further explored as no other member of the 
HC$_{2n}$N was observed in space. In this Letter, we report the 
detection in IRC+10216 of the next member of this family, HC$_4$N, 
and show that the HCCN/HC$_4$N abundance ratio is 20 times smaller 
than the HC$_3$N/HCCN ratio.  

\section{Observations}

Contrary to cyanopolyynes which are linear and have $^1\Sigma$ 
electronic ground states, the lowest energy form of HCCN is a 
quasi-linear triplet (Saito et al. 1984, Rice and Schaeffer 1987). Similarly, 
ab initio calculations \citep{Iku00,Aok93a,Aok93b} predict that one of
the 3 low-energy isomers of HC$_4$N
is a linear triplet with a $^3\Sigma$ ground
state (hereafter HC$_4$N, see Fig. 1), the other two being a C$_3$-ring 
(c$_3$-HC$_4$N) and a bent carbene structure (b-HC$_4$N).  
The 3 isomers have been recently observed in the laboratory 
\citep{Tan99,McC99a,McC99b}; their
microwave spectra are now fully characterized, opening the way 
to unambiguous identification in astronomical sources. 

The astronomical observations were carried out between 1995 and 2003 
using the 30-m IRAM
radiotelescope at Pico Veleta, Sierra Nevada (Spain). Most of the data were
taken during a sensitive 3-mm wavelength (80-115 GHz) line survey of
IRC+10216, which has an r.m.s. noise of few mK per 1
MHz-wide channel. These data were complemented in August 2003 with 
more sensitive observations at the frequencies of SiNC \citep{Gue04}.

We detected a number of U-lines, 9 of which could be grouped into 
3 close triplets with similar line intensities. The center frequencies 
of the triplets were harmonically related, implying a common
linear carrier with a rotation constant 
$B_\circ =2302$ MHz. The rotation constant suggested a molecule 
similar to C$_5$H ($B_\circ= 2287$ MHz) and the triplet structure 
an even number of
electrons. This pointed out to HC$_4$N, and we could quickly check
that the frequencies of our 9 U-lines agreed with those predicted from
the spectroscopic constants of \citet{Tan99}. That HC$_4$N was
the carrier of our new lines was definitely confirmed when we
succeeded in detecting a fourth line triplet (see Fig. 2e and Table 1), during a 
dedicated search made in October 2003.

The observations were made in the wobbling mode, with the secondary  
nutating at a rate of 0.5 Hz, in order to achieve very flat baselines 
(see Fig. 2 and below). Only a zero-order baseline with an offset of
$\simeq 40$ mK, corresponding to the thermal dust emission of the 
circumstellar envelope, has been removed from the spectra.  
Two SIS 3-mm receivers, with orthogonal polarizations and system temperatures
of $100-130\;$K, were used in parallel with two 1.3-mm receivers. The 3-mm
data were co-added, so increasing the effective integration time  
and decreasing the r.m.s. noise. In the case of our best spectrum
(Fig. 2a), the effective integration time and noise 
temperature were equal to 52 h and 0.5 mK per 1-MHz wide channel, 
respectively. 

The temperature scale was calibrated with the help of two absorbers, 
respectively at room and liquid
nitrogen temperature, using the atmosphere transparency model ATM
developed by \citet{Cer85} currently installed in 
the 30-m telescope on-line calibration software (see also Pardo,
Cernicharo and Serabyn 2001).
Pointing and
focus were regularly checked on planets and on the strong nearby quasar
OJ~287. Between the pointing sessions, we monitored the shapes 
of the strong lines that were observed in parallel with the 1.3-mm 
receivers. The cusped shapes and the line horn-to-center intensity 
ratios depend indeed critically on the accuracy of the telescope pointing 
and focus. Thanks to these precautions, the intensities of the lines
observed several times in the 1995-2003 period were found 
constant within 10\%.

\section{Results and Discussion}

Figure 2 shows the spectra covering the 4 triplets, which correspond
to 4 successive rotational transitions of HC$_4$N (from N=$18
\rightarrow 17$ to $21\rightarrow 20$). The
spectrum centered at 82.9 GHz, which was observed while searching for 
SiCN \citep{Gue00}, SiNC \citep{Gue04} and HCCN, is shown twice in this
figure: full scale, i.e. from 82650 MHz to 83150 MHz (Fig. 2a) and 
half-scale, around the HC$_4$N N=18$\rightarrow$17 triplet (Fig. 2b).  
The r.m.s. noise in this 1 MHz resolution spectrum is only 0.5 mK, 
which makes it one of the the most sensitive 
spectra ever obtained with the 30-m telescope. 

Forty spectral lines are detected in Fig. 2a. Most are weak and partly
blended. The blended lines are easily resolved, thanks to their 
characteristic cusps, sharp edges and constant
width in velocity (this shape results from the uniform 
expansion of the outer circumstellar envelope).
We can identify almost 60\% of these lines with the help of the molecular
line catalog maintained
by one of us (JC) and described in \citet{Cer00}. The catalog contains the mm
transitions of some 1200 different molecular species.
Of the remaining 40\% unidentified lines, at least half could be tentatively
assigned to some heavy species (see below).
The carriers of the unambiguously identified lines are mostly $^{12}$C and
$^{13}$C isotopomers of long carbon chain molecules (C$_4$H,
C$_7$H, C$_8$H, H$_2$C$_3$, H$_2$C$_4$, and HC$_5$N), as well as silicon compounds, such
as SiC$_4$, SiCN, and SiNC. For the first time, we detect
a doubly-substituted $^{13}$C isotopomer of cyanoacetylene: H$^{13}$C$^{13}$CCN
(see Fig2e, 96623 MHz; several lines of the three doubly-substituted
$^{13}$C isotopomers of HC$_3$N have
been detected in our 3mm line survey of IRC+10216).

The number of unidentified lines at the 3
mK level detected in the 4 frequency
bands is $\simeq$15 lines/GHz. Several of these lines
probably arise from  
vibrationally excited C$_5$H and C$_6$H and their $^{13}$C isotopomers,
species for which we have no accurate laboratory frequencies
(for example, U82924 and/or U82938 and
U82995 could well correspond to the $J=61/2\rightarrow59/2$ lines
of $^2\Pi_{1/2}$ CCCCC$^{13}$CH; U82677, U82737, U92034 and U92045
could correspond to different $\Sigma$ and $\Pi$ vibronic states
of the lowest energy bending mode of C$_6$H).
Others probably come from $^{13}$C and $^{29}$Si isotopomers of
of SiC$_4$ (for example the J=29-28 and J=32-31 lines
of $^{29}$SiC$_4$ at 87588 and 96648 MHz respectively and
the J=27-26 line of SiC$^{13}$CCC at 82704 MHz).
Finally, some unidentified lines are close to transition frequencies
predicted for SiC$_5$ (U82746, U82876 and U82995), C$_5$S (U83038) and KCN
(for which we have detected several additional lines that
will be published elsewhere), and C$_9$H (U83011).
Obviously, the U-line density is too high at present to claim
the detection of any of these species, which all have a rich mm spectrum,
on the basis of just one, two, or even three weak lines.

The HC$_4$N lines shown in Fig. 2 are cusped (see in particular
Fig. 2a, where the lines have the highest signal-to-noise ratio).
This implies that this molecule is mostly confined into the outer part
of the circumstellar envelope, presumably in the 15"-radius hollow
shell where mos t of the free radicals and carbon chains are found
\citep{Gue93}. From the intensities of the four
triplets we derive a rotational temperature of 25$\pm$4 K, close to
that derived for HC$_3$N. 

The permanent dipole moment of HC$_4$N has been calculated by
\citet{Iku00} to be 4.3 D. Using this value and assuming thermal
equilibrium at 25 K, we derive a column density for HC$_4$N of 1.5
10$^{12}$ cm$^{-2}$.

We have re-calculated the rotation temperature and the abundance of HCCN
by adding to the 2-mm lines reported by \citet{Gue91} the lines
observed in our new 3-mm survey. 
We find a rotational
temperature of 15$\pm$2 K, slightly higher than our previous estimate
(12$\pm$2 K).
The HCCN lines are also cusped and the
HCCN column density, assuming a dipole moment of 3 D \citep{Gue91},
is 1.4 10$^{13}$ cm$^{-2}$.
We note that the increase in
rotation temperature between HCCN and HC$_4$N is not surprising as a
similar increase is observed between the cyanopolyynes HC$_3$N and
HC$_5$N.

The HCCN/HC$_4$N abundance ratio is found equal to 9. This is about twice
the decrement observed between the successive members of the
cyanopolyynes family, HC$_3$N/HC$_5$N, but $20-40$ times less than the 
HC$_3$N/HCCN ($\simeq 200$) and HC$_5$N/HC$_4$N (400) abundance
ratios, which confirms that, like the carbon chain radicals, 
the even and odd HC$_n$N families have parallel, but distinct
formation paths. We have also searched in our 3mm line
survey for the linear triplet HC$_6$N, but without success.
The 3$\sigma$ upper limit to the column density of this species,
assuming a rotation temperature of 25 K, is 1 10$^{13}$ cm$^{-2}$,
not low enough to constrain significantly the HC$_4$N/HC$_6$N ratio, 
but already 
25 times lower than the HC$_7$N column density \citep{Cer87}.

As noted above, the carbon chain molecules C$_{2n+1}$N and
HC$_{2n+1}$N are thought to be formed in IRC+10216 from the 
reaction of radicals such as CCH and CN with
carbon-chain molecules or radicals \citep{Che93,Mil00}.
Similarly, we could consider that HC$_4$N may form through the reaction
of C$_3$N and C$_3$H with HCN and CH$_2$, e.g.,
C$_3$H + HCN $\rightarrow$ HC$_4$N + H, or may be C$_3$N +
CH$_2 \rightarrow $HC$_4$N + H. All the reactants are relatively
abundant in the outer layers of the circumstellar envelope and we expect, 
by analogy with similar reactions studied in the laboratory, that at 
least one of these reactions could proceed rapidly.

Another path proposed for the formation of cyanopolyynes involves
ion-molecule reactions yielding H$_3$C$_n$N$^+$ or H$_2$C$_n$N$^+$, 
followed by the
dissociative recombination of these ions
\citep{Gla86, How90}. The low abundance of ions in IRC+10216 makes
this path relatively slow \citep{Che93}. Moreover, in the case of HCCN and
HC$_4$N,
the reaction of HCN with CH$_3^+$, leading to H$_2$C$_2$N$^+$ +
H$_2$ (or H$_3$C$_2$N$^+$ + H), does not
proceed \citep{Gue91}, making the ion-molecule path even 
slower and less likely than in the case of cyanopolyynes.

Finally, a last process, formation on dust grains followed by
photodesorption when the grains reach the outer envelope layers, 
has been advocated in the case of IRC+10216 \citep{Gue93}. Such 
a mechanism may well produce HCCN, HC$_4$N, and the cyanopolyynes 
observed in the outer shell, but would provide no simple explanation to why
the abundance decrement within each family
seems to be so constant.

The HC$_4$N lines detected during this study belong to the linear
triplet isomer. We have unsuccessfully searched for the singlet bent HC$_4$N 
isomer and for the c$_3$-ring cyclic isomer. Those latter
have a permanent dipole moment almost as large as the linear isomer:
2.96D ($\mu_a$=2.9$\;$D, $\mu_b$=0.5$\;$D) and 3.48$\;$D
($\mu_a$=3.15$\;$D, $\mu_b$=1.49$\;$D),
respectively, versus 4.3 D for linear HC$_4$N \citep{McC99a, McC99b,
Iku00}. The upper limits to the column
density of these species are $\simeq$ 3 10$^{12}$ cm$^{-2}$,
assuming the same rotational temperature than for linear triplet
HC$_4$N (25 K).

The most accurate calculations of the HC$_4$N structure to date (level 
CCSD(T)/cc-pCVTZ) predict that the singlet ring
isomer of HC$_4$N lies 4.2 Kcal/mol below the linear triplet isomer 
\citep{Iku00}. If this result is confirmed, the non-detection of the 
ringed isomer (and of the bent isomer) would further stress the 
discrimination
against ringed structures already noted in the case of
carbon chain radicals.

\begin{table}[htb]
{\bf Table 1: Observed HC$_4$N line parameters}\\
\begin{tabular}{lccr}
\hline
Obs. Freq.  & Calc. Freq. & Transition      & $\int T_A^*dv$    \\
MHz        & MHz      &(N,J$\rightarrow$N',J') &mK.kms$^{-1}$ \\
\hline
82855.6(6) &     82855.52 & 18 17 -17 16   &110(20)\\ 
82884.1(10)$^b$& 82883.57 & 18 18 -17 17   & 94(14)\\ 
82905.4(3)     & 82905.41 & 18 19 -17 18  & 86(10)\\ 
\\
87463.5(10)$^b$ & 87463.15 & 19 18 -18 17     &127(35)\\
87487.9(10) &  87487.87   &  19 19 -18 18     &135(35) \\
87507.6(5) &  87507.31   &19 20 -18 19 	&142(20) \\
\\
92070.0(10)$^b$ &92070.22  &20 19 -19 18 	&90(24) \\
92092.4(10)  & 92092.13    &20 20 -19 19 	&105(24) \\
92109.2(6)   & 92109.48    &20 21 -19 20 	&90(18) \\
\\
96677.1(5)   & 96676.82    &21 20 -20 19 	&75(15) \\
96696.0(10)$^b$ & 96696.32  &21 21 -20 20  &100(23) \\
96712.1(5)    &96711.88     &21 22 -20 21  &75(15) \\
 
\hline
\end{tabular}
\caption{
The number in parenthesis are 1$\sigma$ uncertainties on the last digit,
derived from least square fits; they do not include the 5\% calibration
uncertainty. $^b$ denotes a partly blended line. The frequencies in col. 2
are calculated from the spectroscopic constants measured by \citet{Tan99} in
the laboratory.}
\end{table}

\acknowledgments

We would like to thank Spanish MCyT for funding support for this work
under grants AYA2000-1781, PNIE2001-4716 and AYA2003-02785.

\begin{figure}
\plotone{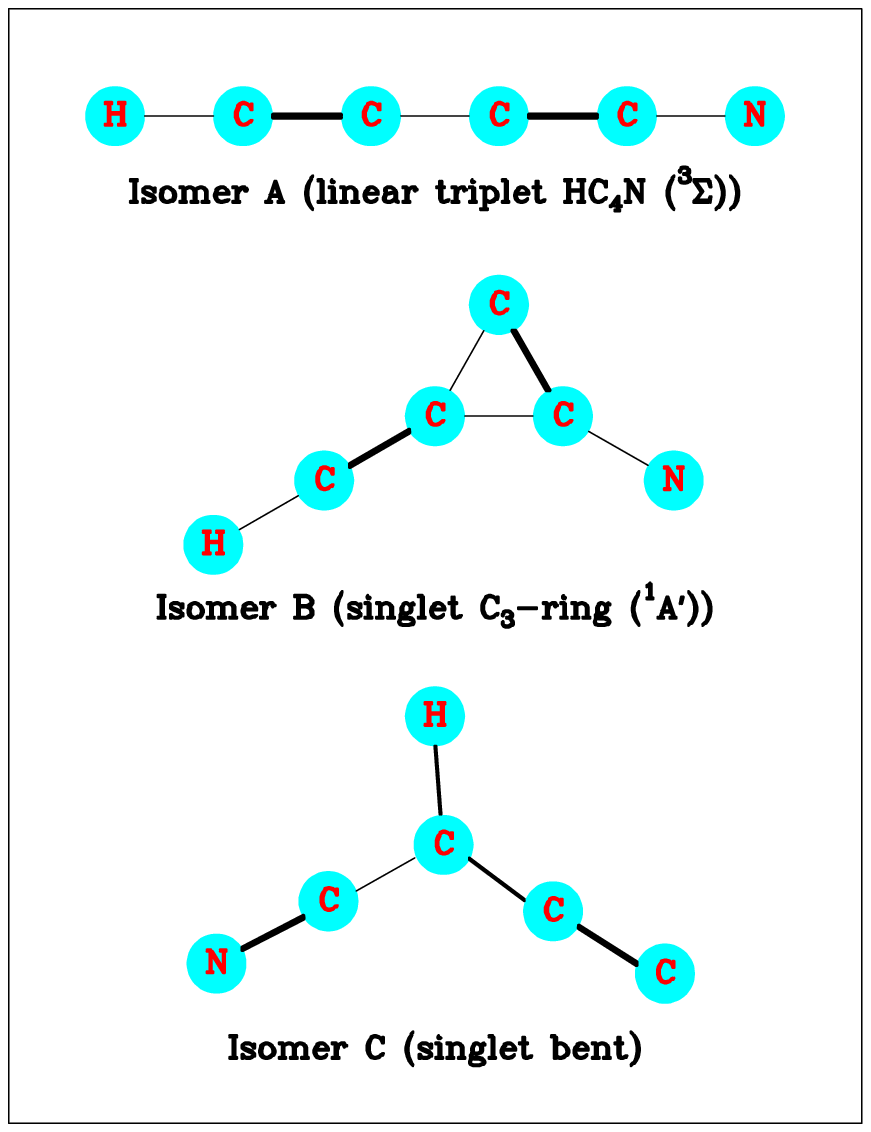}
\caption{The three isomers of HC$_4$N considered in this paper.
The structures are taken from \citet{McC99b} and \citet{Iku00}.
Isomer $A$ is linear
triplet HC$_4$N (HC$_4$N), isomer $B$ corresponds to the C$_3$-ring
structure (c$_3$-HC$_4$N), and isomer $C$ is the singlet bent HC$_4$N
(b-HC$_4$N).
\label{fig1}}
\end{figure}

\clearpage


{\voffset 2.0 cm}

\begin{figure}
\plotone{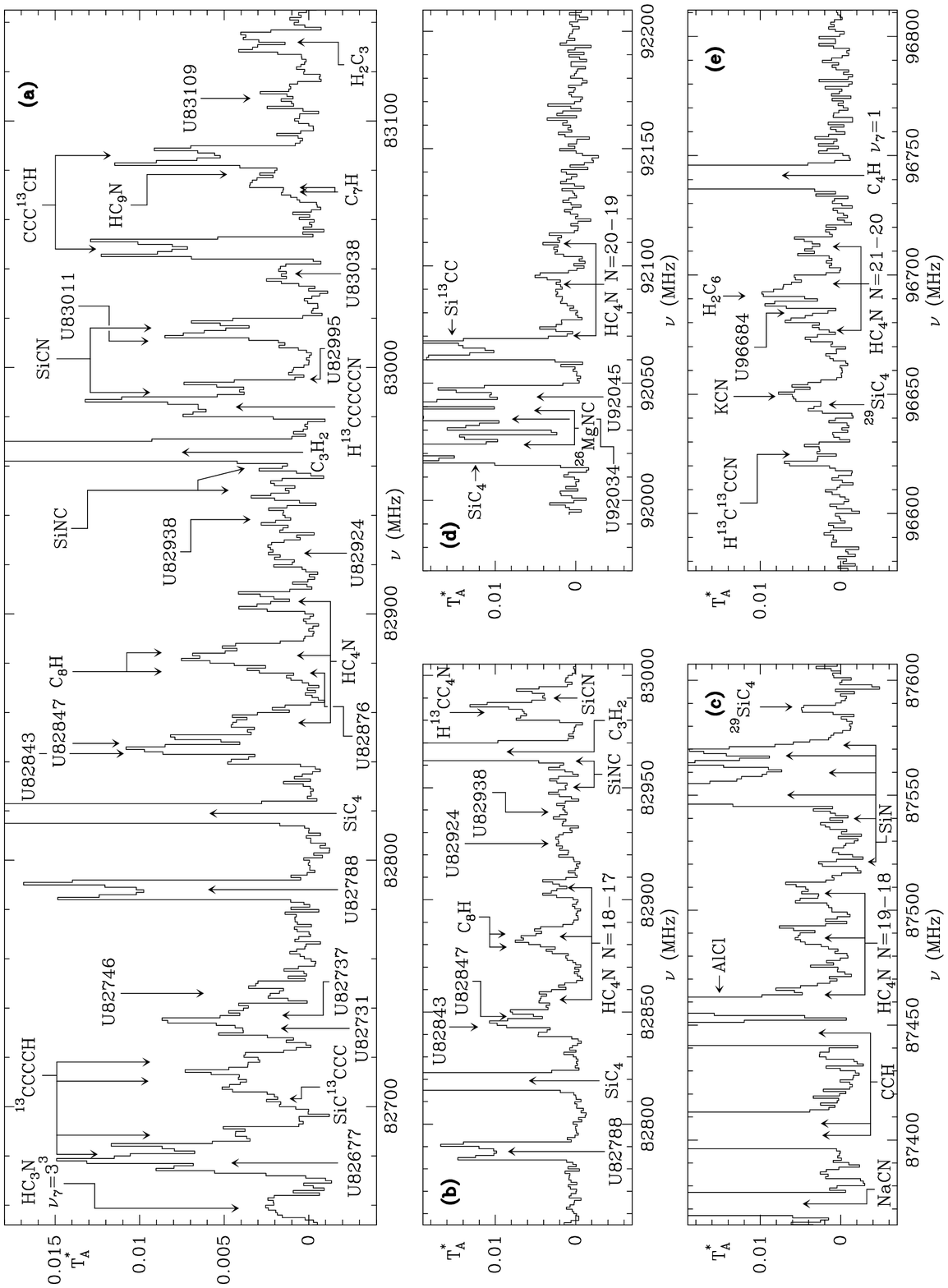}
\caption{Four spectra observed toward IRC+10216 with the IRAM 30-m telescope,
centered on the N=18-17 through 21-20 rotational transitions of
HC$_4$N. The intensity scale is T$_A$$^*$, the spectral resolution 1
MHz. The r.m.s. noise in the upper spectrum is 0.5 mK.  
\label{fig2}}
\end{figure}

\clearpage

\end{document}